\def\cao{\c c\~ao\ }

\def\1{\'{\i}}
\def\beq{\begin{equation}}
\def\eeq{\end{equation}}
\def\bea{\begin{eqnarray}}
\def\eea{\end{eqnarray}}
\def\bed{\begin{displaymath}}
\def\eed{\end{displaymath}}
\def\no{\noindent}

\documentclass[preprint,pre,aps]{revtex4}
\usepackage{graphics,graphicx,amssymb,amsmath,dcolumn,bm,url}

\begin{document}

\title{Effective transport barriers in nontwist systems}

\author{J. D. Szezech Jr.$^1$, I. L. Caldas$^1$, S. R. Lopes$^2$, P.  J. Morrison$^3$, and R. L. Viana$^2$ \footnote{Corresponding author. e-mail: viana@fisica.ufpr.br}}
\affiliation{1. Instituto de F\'{\i}sica, Universidade de S\~ao Paulo, 5315-970, S\~ao Paulo, S\~ao Paulo, Brazil. \\2. Departamento de F\1sica, Universidade Federal do Paran\'a, 81531-990, Curitiba, Paran\'a, Brazil. \\3. Department of Physics and Institute for Fusion Studies\\ The University of Texas at Austin\\ Austin, TX 78712 USA.}

\begin{abstract}
In fluids and plasmas with zonal flow reversed shear,  a peculiar kind of transport barrier appears in the shearless region, one that is associated with  a proper route of transition to chaos. Using a symplectic nontwist maps, which model such  zonal flows,  we  investigate these barriers. In particular,   the standard nontwist map, a paradigm of nontwist systems, is used to  analyze the parameter dependence of the transport through a broken shearless barrier. Varying a proper control parameter, we identify the onset of structures  with high stickiness that give rise to an effective barrier near the broken shearless curve.  Several diagnostics are used:  stickiness is measured by escape time and transmissivity plots, while barriers  that separate different regions of stickiness are   identified by employing  two indicators in the phase space, the finite-time rotation number and the finite-time Lyapunov exponent. 
\end{abstract}
\date{\today}

\maketitle

\section{Introduction}
\label{intro}

Internal transport barriers  that appear in Hamiltonian dynamical systems have been proposed as an explanation for  the  cessation or reduction of  transport  in physical systems  that describe  fluids (e.g.\ \cite{behringer,castillo93}) and plasmas (e.g.\ \cite{naulin}). 
These barriers may have various  physical or dynamical origins, yet they can and have been used to control experiments  and sometimes to improve desired confinement of trajectories.  Thus,  there is justification for studying these barriers in the general context of Hamiltonian systems, which we do here. 


A peculiar kind of transport barrier exists in fluids and plasmas with a nonmonotonic equilibrium zonal flow, which give rise to  orbit topologies that can only exist with reversed shear \cite{castillo00,castillo92}, i.e., with a  nonmonotonic rotation number profile. The barriers appear in the shearless region of nontwist Hamiltonian dynamical systems and present their own typical characteristics with a proper route of transition to chaos \cite{castillo96}.  They possess robustness -- persisting even for high amplitude perturbations -- and have an effective capacity to reduce the transport even after invariant tori are  broken \cite{jefferson}.    Invariant barriers persist until the destruction of the shearless invariant curve \cite{castillo93,castillo00}, but the capacity to reduce transport remains and is  credited to the stickiness around islands that remain  in the shearless region \cite{szezech09}.

Such barriers have been numerically and experimentally identified in several nontwist dynamical systems such as those that describe magnetic field lines in toroidal plasma devices with reversed magnetic shear \cite{morrison00,kroetz}, the advection of a passive scalar by an incompressible shear flow \cite{pierrehumbert}, travelling waves in geophysical zonal flows \cite{behringer,castillo92,castillo93,castillo00}, the ${\bf E}\times{\bf B}$-drift motion of charged particles in a magnetized plasma under the action of a time-periodic electric field from an electrostatic wave \cite{castillo00,horton98,marcus08}, and laser-plasma coupling \cite{langdon}. 

The mentioned barrier properties have been theoretically derived for the standard nontwist map (SNTM), a paradigmatic example of nontwist systems proposed by del Castillo-Negrete and Morrison in 1993 \cite{castillo93}, and interpreted as a consequence of successive bifurcations of the shearless invariant curve \cite{castillo00,castillo96,wurm05}. This scenario shows for the nontwist standard map the relevance of the location of the shearless region where the transport reduction occurs. Thus, for all nontwist systems, the transport reduction should be observed in the shearless region and not necessarily in high shear regions as for other barriers proposed to exist  in twist systems \cite{terry}.

The standard nontwist map can be regarded as arising from intersections of phase space trajectories with a given surface of section. They can also appear as stroboscopic samplings of trajectories in a time-dependent system at fixed time intervals. This canonical map is convenient for investigating transport, inasmuch we can compute a large number of iterations in a short time with minimal propagation of numerical error, as occurs with most schemes for solving differential equations. This property is particularly important in studies of Hamiltonian transport, which require computation of phase space trajectories over very long time intervals.

In nontwist systems, after the shearless curve breakdown,   chaotic orbit stickiness is high in the shearless region and, consequently, the chaotic transport is reduced in this region. For the standard nontwist map, this local transport reduction has been associated with an effective transport barrier and characterized in terms of the orbit escape time and transmissivity \cite{szezech09}. Moreover, it was suggested that the sensitive dependence of these quantities has the same parameter dependence as the dominant crossings of stable and unstable manifolds \cite{szezech09}.

In this paper, for some parameter ranges for which the transport barrier of the standard nontwist map \cite{castillo93} is broken, we identify remaining stickiness structures that reduce the transport in the phase space region through the broken shearless barrier. We show how these stickiness structures, determined by the homoclinic tangles of the dominant remaining dimerized islands, change with the control parameters and modify the observed transport within the shearless region. Moreover, we also identify these structures, in phase space,  by mapping out the finite time Lyapunov exponent (FTLE) \cite{yuan} and the finite time rotation number (FTRN) \cite{nosso} in the shearless region.  The FTLE and FTRN  allow us to identify structures that separate different regions of stickiness in phase space. Other indicators have been used in Hamiltonian systems to delineate regions with other dynamical properties, besides the stickiness, as  Lagrangian coherent structures \cite{haller04} and resonant zones \cite{easton}, but we note the convenience and simplicity of the FTLE for identifying the stickiness structures and effective barrier onset. 

The rest of the paper is organized as follows: in Section \ref{sntm} we describe two qualitatively different transport regimes related to the separatrix reconnection and breakup of shearless curves in the SNTM. Section \ref{tranbar} is devoted to a characterization of the transport barriers, which are stickiness structures of the SNTM. Section \ref{rotnum} uses the infinite-time rotation number to evidence period-three satellite islands just after the formation of transport barriers. Section \ref{indic} introduces the finite-time rotation number, an indicator for stickiness structures. Section \ref{RNridges} uses the finite-time rotation number ridges to visualize transport-related escape channels. Finally Section \ref{conclu} contains our Conclusions.


\section{Standard nontwist map}
\label{sntm}

We  consider two-dimensional area preserving maps of the following general form:
\bea
\label{geraly}
y_{n+1} & = & y_n + f(x_n), \\
\label{geralx}
x_{n+1} & = & x_n - g(y_{n+1}), \quad (mod \, \, 1) 
\eea
where $(x_n,y_n)$ are the normalized angle and action variables, respectively, of a phase space trajectory at its $n$th piercing with a Poincar\'e surface of section. We assume that $y_n \in \mathbb{R}$ and $x_n \in (-1/2,1/2]$, and also that $f(x_n)$ is a period-$1$ function of its argument, representing a perturbation strength (whose time-dependence is a periodic delta function). 

The function $g(y_{n+1})$ is the winding number of the unperturbed phase-space trajectories lying on nested tori, its derivative being the so-called shear function. If the function $g$ is monotonically increasing or decreasing for all values of interest, the corresponding shear does not change sign, which amounts to the following twist condition:
\beq
\label{twistcond}
|g'(y_{n+1})| = \left\vert \frac{\partial x_{n+1}}{\partial y_n} \right\vert \ge c > 0,
\eeq
where $c \in \mathbb{R}$. The loci where $g'(y_S) = 0$, i.e.\  where the shear changes sign, define shearless curves in phase space.

We focus on systems for which the winding number $g(y)$ in a  representative map of the form (\ref{geraly}) and (\ref{geralx}) violates (\ref{twistcond}) by having one extremum at $y = y_S$, at which the shear changes sign. In the vicinity of this point the lowest-order approximation for the function $g(y)$ is a quadratic one. If, in addition, we keep only one sinusoidal mode in the perturbation term of (\ref{geraly}) and (\ref{geralx}), we obtain the standard nontwist map (SNTM), introduced in Ref. \cite{castillo93}:
\begin{eqnarray}
\label{sntmx}
x_{n+1} & = & x_n + a(1-y_{n+1}^2),\\
\label{sntmy}
y_{n+1} & = & y_n - b \sin(2\pi x_n)\ ,
\end{eqnarray}
where $x \in [-1/2,+1/2) $, $y \in {\mathbb R}$, $a \in (0,1)$, and $b > 0$. 

Let us first consider the unperturbed case ($b=0$). The twist condition (\ref{twistcond}) is violated at the point $y_S = 0$, defining a shearless curve $\{ (x,y) | -1/2 < x \le 1/2, y = y_S = 0 \}$. The quadratic form of $g$ around $y_S = 0$ leads to two invariant curves, at $y = \pm y_0$ with the same winding number $a(1-y_0^2)$ at both sides of the shearless curve. As the perturbation is switched on ($b \ne 0$) two periodic island chains appear at the two invariant curve locations, and the former shearless curve becomes a shearless invariant tori separating these two island chains. There are also chaotic layers attached to the ``separatrices'' of both island chains, as expected from the presence of homoclinic crossings therein. These chaotic layers are not connected, though, as far as there are invariant curves near the shearless invariant tori acting as dikes, preventing global transport.

\begin{figure}
\includegraphics[width=0.7\columnwidth,clip]{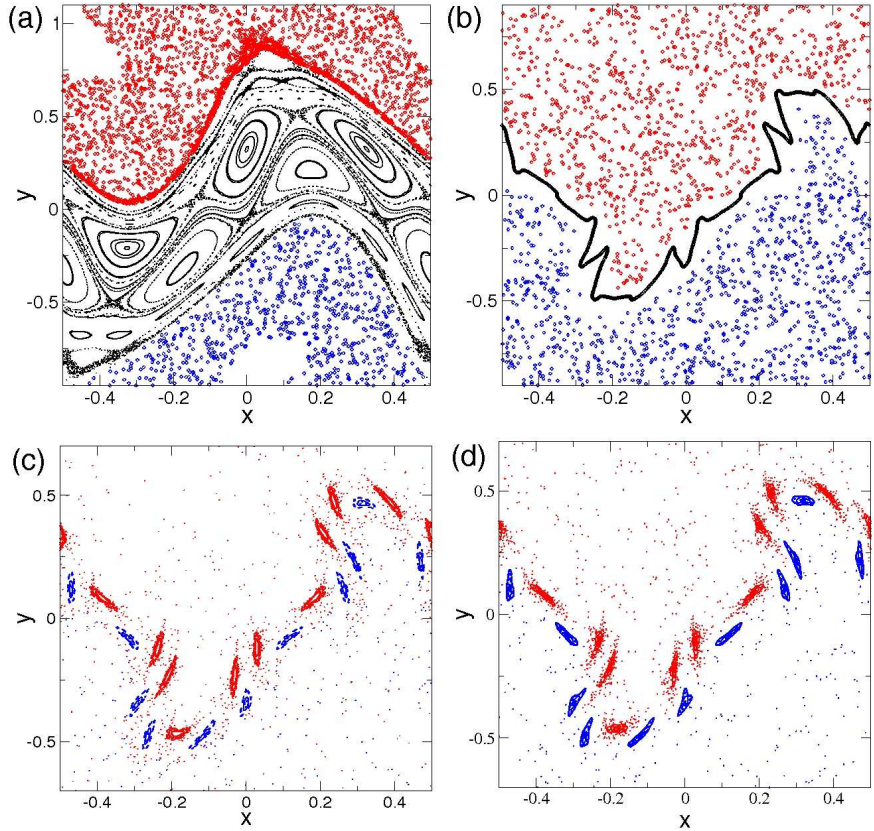}
\caption{\label{phase} (color online) Poincar\'e section of the standard nontwist map (\ref{sntmx}) and (\ref{sntmy}) for $b=0.6$ and (a) $a = 0.364$; (b) $a = 0.8$; (c) $a = 0.80552$ and (d) $a = 0.8063$.}
\end{figure}

A representative example is depicted in  Fig.~\ref{phase}(a), where a Poincar\'e section of the SNTM is depicted for $b = 0.6$ and $a = 0.364$ (in the following we shall fix this value of $b$ and vary only the parameter $a$). We observe two island chains with three islands each, with winding number $0.3$. In the unperturbed map the corresponding invariant curves are located at $y_0 = \pm 0.42$. The local maxima of the perturbed winding number profile define a shearless invariant curve, whose existence can be inferred between the two island chains. The island chains bordering the shearless invariant curve are transport barriers, since chaotic trajectories above and below do not mix at all (in Fig.~\ref{phase}(a) they have been represented in different colors). 

If the parameters are further modified another noteworthy feature of nontwist maps can emerge, depending on the parameter space region. In one scenario (generic reconnection) the island chains with the same winding number approach each other and their unstable and stable invariant manifolds suffer reconnection. In the region between the chains, there   appear new invariant tori called meandering curves (which are not KAM tori, though, since the latter must be a graph over $x$, while meanders are not). The periodic orbits remaining eventually coalesce and disappear, leaving only meanders and the shearless torus. This set is a robust transport barrier, as illustrated in  Fig.~\ref{phase}(b), where two chaotic orbits on different sides of the barrier are kept segregated by a shearless curve \cite{castillo96}. The other possible reconnection scenario is nongeneric and involves the formation of vortex pairs, which is only possible in nontwist maps with symmetries \cite{wurm05}. 

Further growth of the $a$-parameter causes the breakup of the transport barrier and the consequent mixing of the chaotic orbits formerly segregated on both sides of the shearless invariant torus [Figs.~\ref{phase}(c) and (d)]. However, there are subtle differences between Figures \ref{phase}(c) and (d), which are ultimately related to the invariant manifold of unstable periodic orbits (saddles) embedded in the chaotic orbit. 

In order to highlight these differences we resort to numerical diagnostics of the transport properties along the globally chaotic layer. The first diagnostic, the escape time, is computed as follows: for fixed values of the parameters $(a,b)$ we consider a large number of initial conditions in a grid of $N_P = 2 \times 10^6$ points regularly spaced in the square $[-0.5,0.5] \times [-0.9,0.9]$. Each initial condition was iterated until the ensuing trajectory crosses either one of two boundaries placed at $(-0.5 < x_B < 0.5,y_B = \pm 2.0)$. The escape time $E$ is the time it takes for a given trajectory to reach any of these boundaries and, since it varies widely with the initial condition, we work with an average escape time ${\bar E}$. 

Another diagnostic of transport is the transmissivity, which is the fraction of orbits crossing the (formerly existent) transport barrier. It is computed by randomly choosing a large number $N_P = 4.5 \times 10^6$ of initial condition on the line $(-0.5 < x_B < 0.5,y = 1.0)$ and iterated the resulting orbits for $T = 2.0 \times 10^6$ times. Then we count the number of orbits that cross the broken barrier reaching the $y=-1$ line. The difference between escape time and transmissivity is that the former considers orbits escaping from the region near the transport barrier through both sides of it, whereas the latter only takes into account those orbits that actually cross the transport barrier. 

\begin{figure}
\includegraphics[width=0.9\columnwidth,clip]{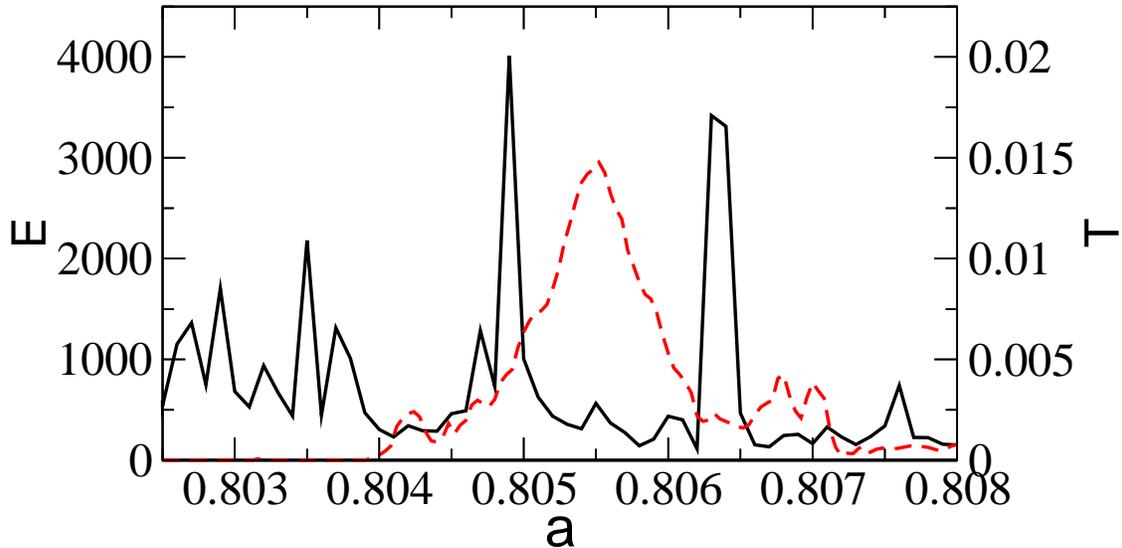}
\caption{\label{escapefig} (color online) Average escape times $E$ (black full line) and transmissivity $T$ (red dashed line) in the standard nontwist map of (\ref{sntmx}) and (\ref{sntmy}) for $b=0.6$ and $a$ values in the vicinity of the shearless curve breakup.}
\end{figure}

In Fig.~\ref{escapefig} we depict the average escape time ${\bar E}$ and the transmissivity $T$ as $a$ sweeps through an interval of values in the vicinity of the ``last" shearless curve breakup. The transmissivity (red dashed line) becomes nonzero only after this shearless curve has been broken up [such as in Fig.~\ref{phase}(c) and (d)], whereas the escape time is finite even before this breakup [Fig.~\ref{phase}(b)]. For the parameter value used in the Poincar\'e section depicted in Fig.~\ref{phase}(c) the transmissivity is quite high (the large red peak in the middle of Fig.~\ref{escapefig}), with the escape time being relatively low. By way of contrast, in the seemingly identical Fig.~\ref{phase}(d) the transmissivity is dramatically reduced (from $\sim 6000$ to {\it circa} $1000$), while the escape times remain practically the same. This suggests   there is a mechanism   preventing orbit escape through the former barrier, nevertheless allowing diffusion at both sides of the barrier. Hence, even though the last shearless curve has been broken an effective barrier can remain, which we explore next.

\section{Transport barrier}
\label{tranbar}

\begin{figure}
\includegraphics[width=0.5\columnwidth,clip]{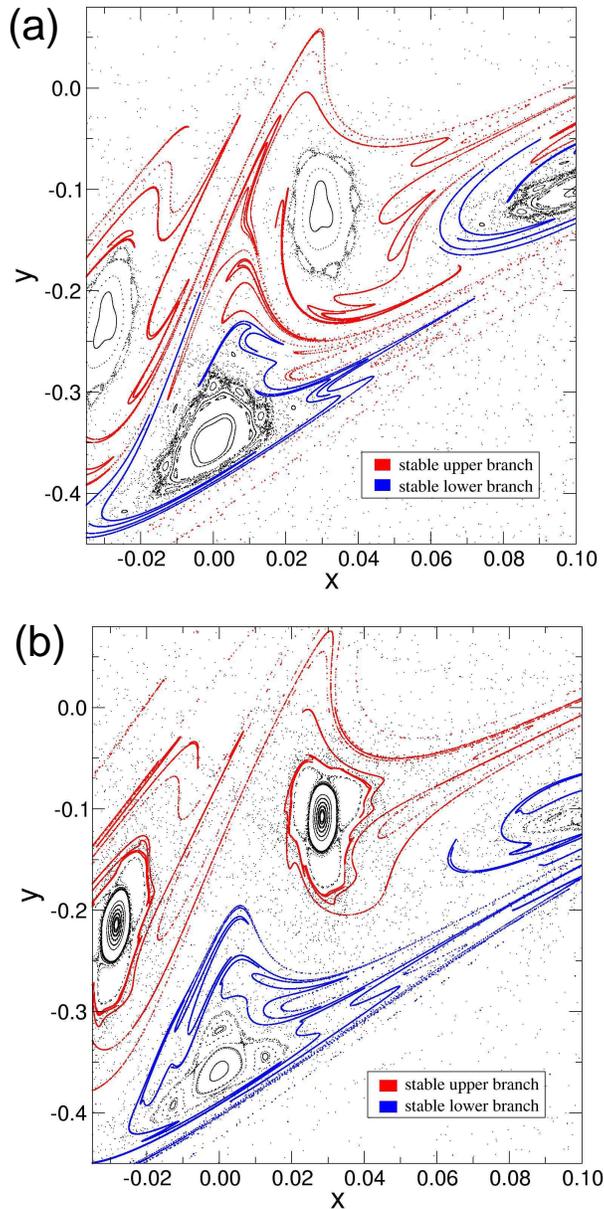}
\caption{\label{manifig} (color online) Invariant stable manifolds of periodic orbits embedded in the chaotic region after the breakup of the last shearless curve for the standard nontwist map with $b = 0.6$ and (a) $a = 0.80552$ and (b) $a = 0.8063$. The red and blue curves represent  manifolds of the upper and lower period-$11$ island chain remnants.}
\end{figure}

This effective transport barrier is a consequence of a topological reordering of the invariant stable and unstable manifolds of periodic orbits embedded in the chaotic region that follows the breakup of the last shearless curve \cite{corso}. This chaotic region coexists with the remnants of   ``twin" period-$11$ island chains \cite{szezech09}. The stable and unstable manifolds of the saddle points in the chaotic region therein suffer a reconnection at a value of $a$ between those used to obtain Figs.~\ref{phase}(c) and (d). Before this reconnection,  the manifolds belonging to the upper and lower period-$11$ island chains cross each other many times (intercrossing), forming an escape channel responsible for the high transmissivity displayed by Fig.~\ref{phase}(c), as illustrated by Fig.~\ref{manifig}(a), where the stable manifolds of the upper and lower islands are depicted in red and blue, respectively. 

On the other hand, in Fig.~\ref{manifig}(b) [which corresponds to the Poincar\'e section  of Fig.~\ref{phase}(d)] the manifolds of the upper (lower) island chain have chiefly crossings with manifolds of the upper (lower) chain (intracrossings), hence diminishing transmissivity, while still allowing for diffusion. The changing manifold structure responsible for the local decrease of cross-barrier diffusion determines the local stickiness, the characterization of which is the purpose of the Sections \ref{rotnum} and \ref{RNridges}.

\begin{figure}
\includegraphics[width=0.5\columnwidth,clip]{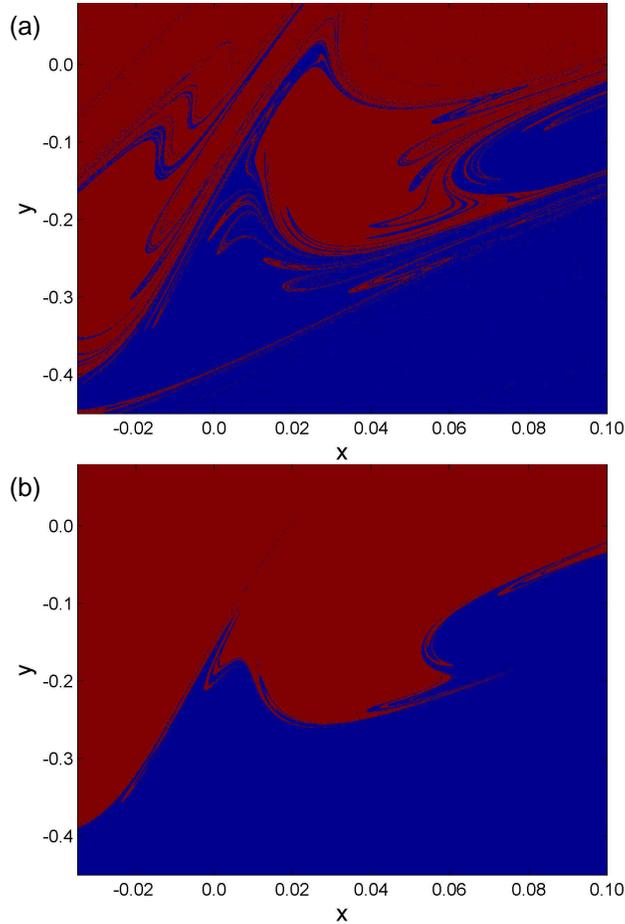}
\caption{\label{transfig} (color online) Fixed-time average $y$-values for orbits evolving from a grid of initial conditions  under  the standard nontwist map, with $b = 0.6$ and (a) $a = 0.80552$ and (b) $a = 0.8063$. If $<y>$ is positive (negative) the corresponding initial condition is plotted in red (blue).}
\end{figure}

Another illustration of the manifold reconnection forming transmissivity channels is provided by the numerical experiment depicted in Fig.~\ref{transfig}. Here we considered a grid of initial conditions and computed for each point the average $y$-value for a given time $t_{esc} = 100$. If $<y>$ was positive (negative) the corresponding initial condition was plotted in red (blue). If there were a perfect transport barrier, like a shearless curve between the upper and lower twin chains, then  there would be a clearcut separation between points evolving to positive large $y$-values (red) and negative large values of $y$ (blue). After the breakup of the last shearless curve, the case of high transmissivity [Fig.~\ref{transfig}(a)] clearly shows the existence of incursive fingers of the blue region, showing that there are initial conditions above the upper chain going to negative $y$ through the blue channels. The low-transmissivity [Fig.~\ref{transfig}(b)] suggests that the manifolds after reconnection act as effective transport barriers, with very small diffusion between the colors and few identified incursive fingers.  In the next Section we consider the FTRN as a means for diagnosing this situation. 

\section{Rotation number}
\label{rotnum}

Let $x \mapsto M(x)$ be a map of the circle ${S}^1$ onto itself. If the dynamical system is a continuous-time flow, then $M$ can be thought of as a Poincar\'e map obtained through successive intersections of the trajectories with a given surface of section in the phase space. The rotation number for the trajectory starting at the point $x_0$ is defined as 
\beq
\label{rotation}
\omega = \lim_{n\rightarrow\infty} \Pi\cdot({M^n(\mathbf{x}_0) - \mathbf{x}_0})/{n},
\eeq
\no which is lifted to $\mathbb{R}$ and $\Pi$ is a suitable angular projection. According to a theorem of Poincar\'e, if $f$ is orientation-preserving this limit exists for every initial condition $x_0 \in {S}^1$ and does not depend on $x_0$ as well \cite{katok}. 

As a simple example, let us consider a rigid rotation on the circle ${S}^1$ given by $M(x) = x + w$. The corresponding rotation number is $\omega = w$. If $w$ is a rational number $p/q$, where $p$ and $q$ are coprime integers, the trajectory represents a period-$q$ orbit of the map $M$, and $p$ is the integer number of times the orbit cycles through the $x$-direction before returning to its initial position. If $w$ is irrational, then the ensuing (quasiperiodic) orbit densely covers the circle ${S}^1$. The rotation number is not defined for chaotic orbits, for the limit in (\ref{rotation}) does not exist in general.

\begin{figure}
\includegraphics[width=1.0\columnwidth,clip]{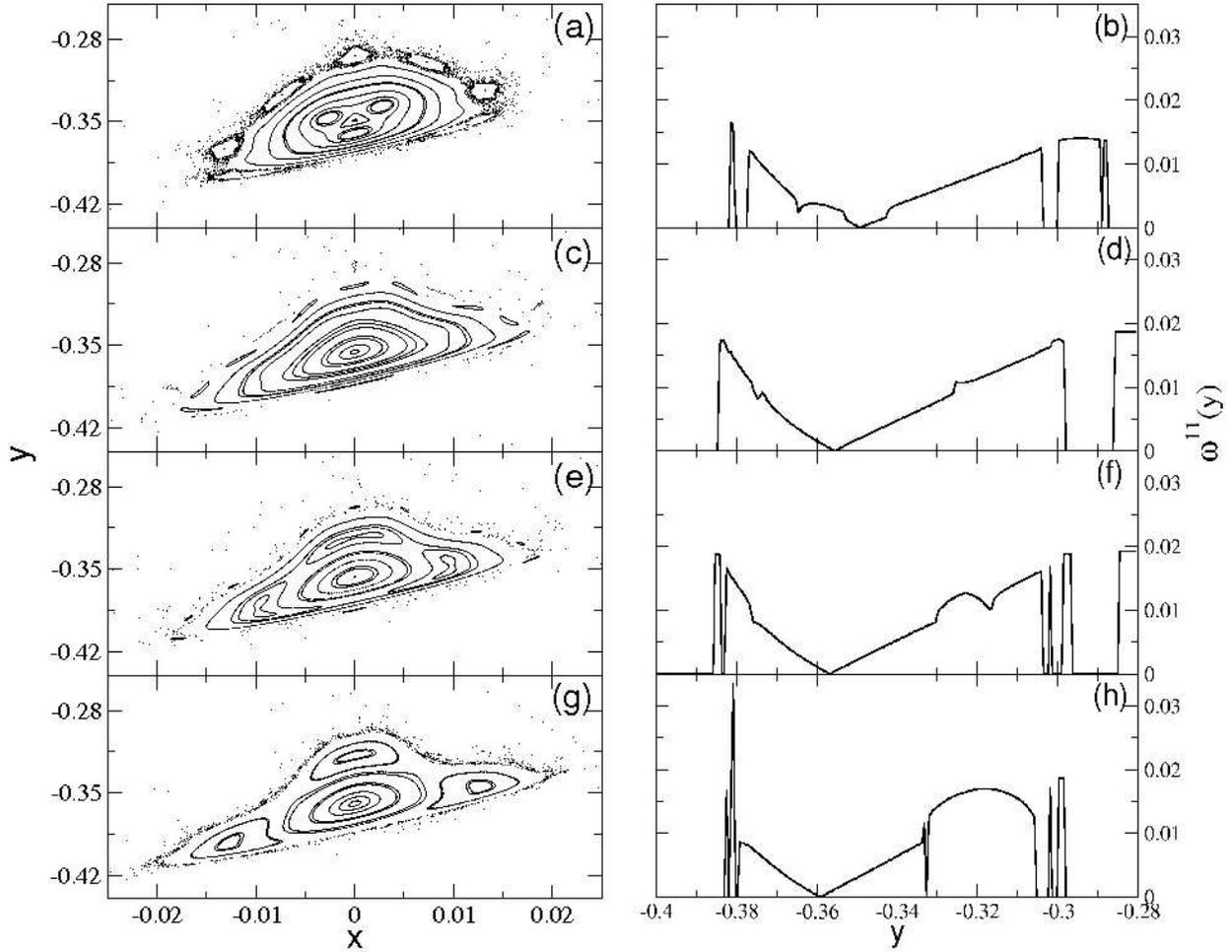}
\caption{\label{rotfig} Left: one island of the period-$11$ primary chain in the chaotic region after the breakup of the last shearless curve for the standard nontwist map with $b = 0.6$ and (a) $a = 0.80552$; (c) $a = 0.8060$; (e) $a = 0.8061$; (g) $a = 0.8063$. Right: corresponding local rotation number profiles at $x = 0$ cross-sections.}
\end{figure}

The rotation number profiles of the period-$11$ island chain yield information about the topological mechanism underlying the manifold reconnection that we  described in Section \ref{tranbar}, and which creates an effective transport barrier. For this sake we have considered a single island of this chain [left panels in Fig.~\ref{rotfig}] and the local rotation number profiles corresponding to cross sections of them taken at $x = 0$ [right panels in Fig.~\ref{rotfig}]. Each island of the primary period-$11$ chain of the high transmissivity case [Fig.~\ref{rotfig}(a)] is characterized by having an outer secondary period-$7$ chain and an inner secondary period-$3$ chain. 

\begin{figure}
\includegraphics[width=0.8\columnwidth,clip]{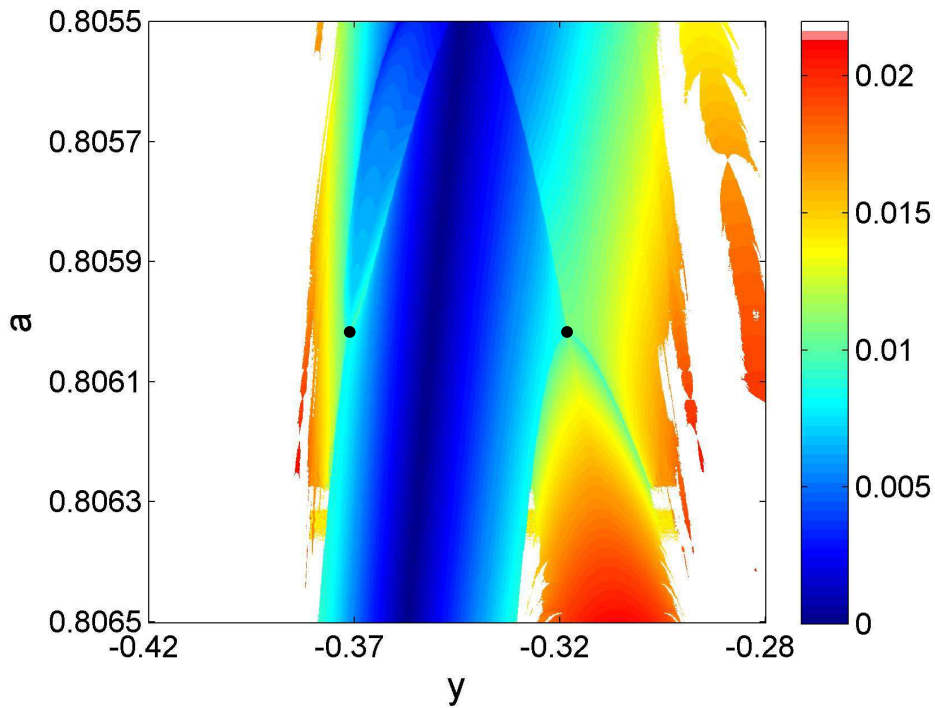}
\caption{\label{parfig} (color online) Rotation number (in colorscale) as a function of $y$ for $x = 0$ cross-sections and different values of $a$ for the standard nontwist map, keeping $b = 0.6$. White pixels stand for values for which the rotation number is not well-defined, since the corresponding orbit is chaotic.}
\end{figure}

As we approach the point of manifold reconnection [Fig.~\ref{rotfig}(c)] the period-$7$  chain is engulfed by the surrounding chaotic sea, and an outer period-$10$ chain emerges. The inner period-$3$ chain, however, seems to disappear. Its reappearance [Fig.~\ref{rotfig}(e)] occurs slightly after the transport barrier is formed, being also present in the low-transmissivity situation [Fig.~\ref{rotfig}(g)]. This transition appears in Fig.~\ref{parfig} marked by the two black dots that indicate the period-$3$ chain death and birth, for a critical parameter $a$, in different $y$-values.

The topological changes occurring while the manifolds reconnect can be also appreciated from the point of view of the rotation number profiles in the right panels of Fig.~\ref{rotfig} or, alternatively, by the diagram depicted in Fig.~\ref{parfig}, where the values of the rotation number are shown in colorscale as a function of $y$ for continuous variation of $a$-values. As a common trend, the formation of transport barrier through manifold reconnection is followed by increasingly high values of the rotation number. This fact suggests  that the transport barrier itself may be somewhat connected with comparatively large values of the rotation number, and this suggests  a diagnostic based on the rotation number, which we consider in Section \ref{indic}.   We also emphasize the appearance of a period-$3$ satellite island (``gumdrops'') just after the formation of the effective transport barrier.

\section{Indicators}
\label{indic}

From  the example worked out in Sections \ref{sntm}-\ref{rotnum}, we learned there is an effective transport barrier related to manifold reconnection generating intracrossings,  and thus the large expansion rates are related to the escape routes avoiding the barrier. 
From the observations of Section \ref{rotnum}  the barrier formation is seen to be accompanied by a localized increase in the rotation number -- this suggests  that the barrier can be related to large values of finite-time approximations of the local rotation number.  In a previous work \cite{nosso} we presented the  same idea for identifying coherent structures.   The FTRN is a  simpler and computationally faster method than the FTLE for one and a half degree-of-freedom systems, because the FTRN does not require the evaluation of spatial derivatives or additional differential equations and this  substantially reduces the computer time.  We will pursue both indicators here to determine barriers. 

\begin{figure}
\includegraphics[width=0.7\columnwidth,clip]{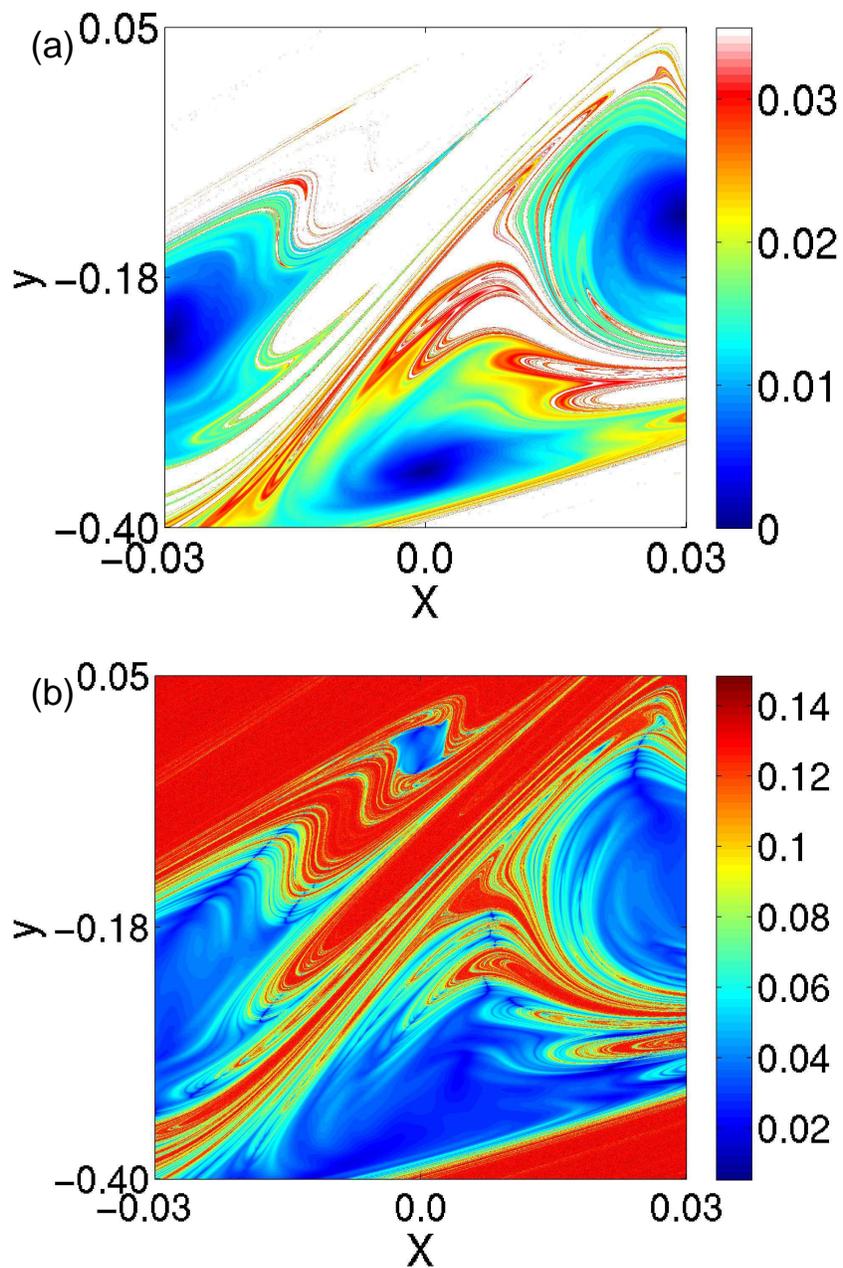}
\caption{\label{finifig1} (color online) (a) Finite-time rotation number and (b) finite-time largest Lyapunov exponent (in colorscale) as a function of the initial condition for the standard nontwist map, with $a = 0.80552$ and $b = 0.6$.}
\end{figure}

The time-$N$ finite-time rotation number (FTRN) is thus the time-$N$ truncation for the corresponding iterations of the map $M$
\beq
\label{ftrn}
\omega_N(\mathbf{x}_0):=  \Pi\cdot \frac{M^N(\mathbf{x}_0) - \mathbf{x}_0)}{N},
\eeq
\no In general, $\omega_N$, like any truncation, depends on the initial condition. While the infinite-time rotation number is not well-defined for chaotic orbits, its finite-time counterpart exists for any orbit, chaotic or not. The FTRN measures the average rotation angle swept by a trajectory over a time interval $T$, and thus conveys information about the local behavior of trajectories, in the same way as the finite-time Lyapunov exponents does (since the latter is the  local rate of contraction or expansion). 

In Fig.~\ref{finifig1}(a) and (b) we show the FTRN and largest FTLE respectively, of a region near the period-$11$ island chain in the high transmissivity case previously shown in the manifold diagram of 
Fig.~\ref{manifig}(a). The transport channel provided by the intercrossing of manifolds of the upper and lower chains is illustrated by the striations of constant rotation number or Lyapunov exponent crossing the phase space channels between the islands. 

\begin{figure}
\includegraphics[width=0.7\columnwidth,clip]{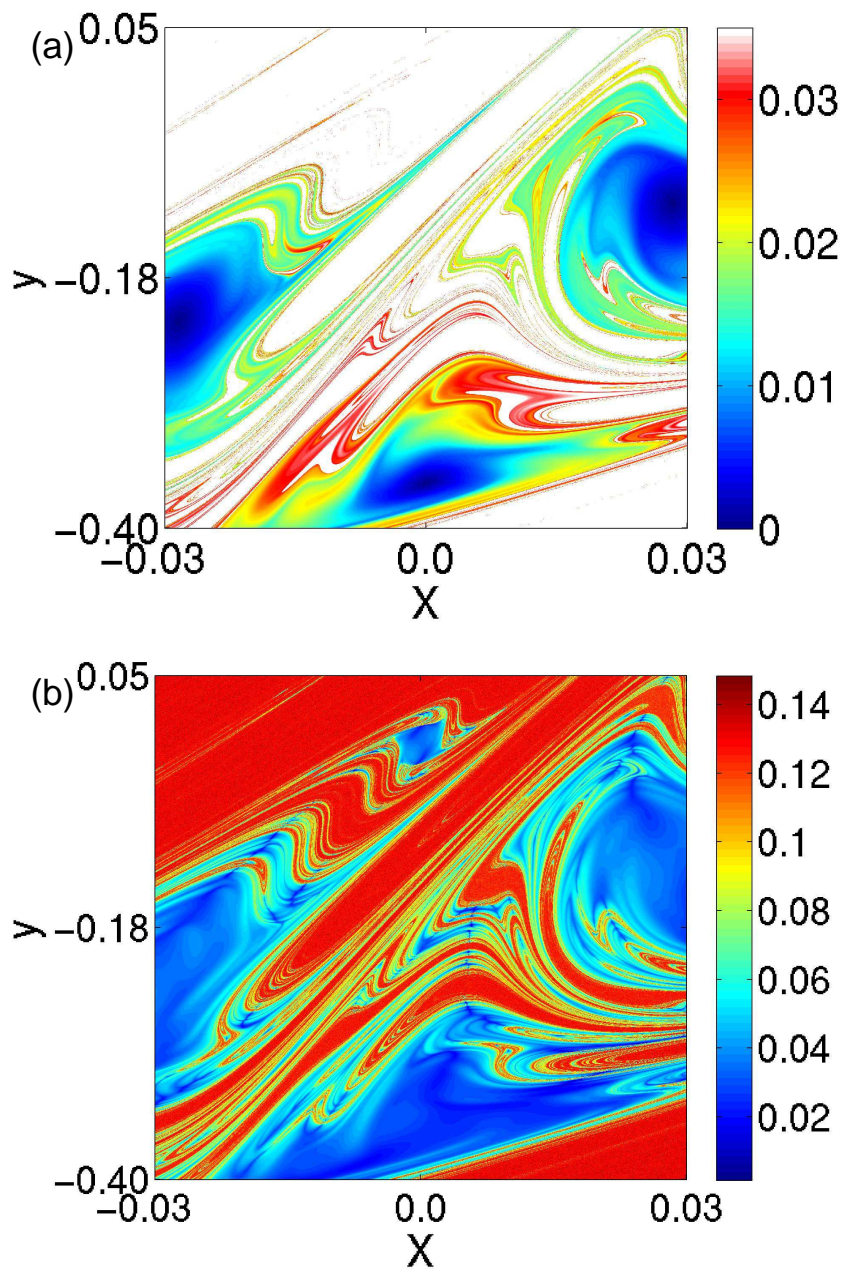}
\caption{\label{finifig2} (color online) (a) Finite-time rotation number and (b) finite-time largest Lyapunov exponent (in colorscale) as a function of the initial condition for the standard nontwist map, with $a = 0.8063$ and $b = 0.6$.}
\end{figure}

The FTRN and FTLE plots corresponding to the low-transmissivity case are depicted in Fig.~\ref{finifig2}(a) and (b), respectively. 
Thus we see that the FTRN is an indicator of effective transport barriers and stickiness structures that survive in the chaotic region after the invariant curves are broken. These transport characteristics are essentially related to the stickiness and recurrence and have been also described with the FTLE \cite{grasso} for magnetic reconnection. Here we see that the same characteristics can be more easily obtained by calculating the FTRN, since comparable results with  less computer time give essentially the same picture.  Therefore,   the FTRN is a ``fast indicator'' of barriers. 

On the other hand, the FTRN (as well as the FTLE) are not  good chaos indicators.  In the infinite time limit the Lyapunov exponent can be used to distinguish chaotic and regular trajectories, but  it is not efficient, and  the rotation number is defined only for the regular (periodic or quasi-periodic) trajectories. However, as seen here a grid of both FTRNs and FTLEs in the two-dimensional phase space do reveal structures  coincident with regions of  observed stickiness (around resonances) and effective barriers that are not detected for long time observations.  We argue in favor of  the FTRN,   since it is  faster with the FTLE, and proceed with its further use in the next section.


\section{Ridges of finite-time rotation number}
\label{RNridges}

FTRNs also provide a convenient way to visualize the escape channels related to inter and intracrossing transport. In order to do that, we plot the ridges associated with a field of FTRNs, representing crests of local maxima. This is defined as follows:  suppose   one has computed the FTRN field in a two-dimensional region $\omega_N(x,y)$.  Then, according to a procedure developed by Marsden et al.\  \cite{marsden} we superimpose an $N \times N$ mesh of equally spaced initial conditions $(x_i,y_j)$, with $i,j=1, 2, \ldots N$ and compute the related Hessian matrix
\beq
\label{hesse}
{\bf H}(x,y) = \left(
\begin{array}{cc}
\frac{\partial^2 \omega_N(x,y)}{\partial x^2} & \frac{\partial^2 \omega_N(x,y)}{\partial y \partial x} \\
\frac{\partial^2 \omega_N(x,y)}{\partial x \partial y} & \frac{\partial^2 \omega_N(x,y)}{\partial y^2}
\end{array}
\right) = 
\left(
\begin{array}{cc}
\omega_{xx} & \omega_{xy} \\
\omega_{xy} & \omega_{yy} 
\end{array}
\right),
\eeq
\no where the corresponding derivatives must be computed for all mesh points, e.g. 
\beq
\label{hesseder}
\frac{\partial \omega_N(x,y)}{\partial x} = \left(
\begin{array}{cccc}
\frac{\partial \omega_N(x_1,y_1)}{\partial x} & \frac{\partial \omega_N(x_1,y_2)}{\partial x} & \cdots & \frac{\partial \omega_N(x_1,y_N)}{\partial x} \\
\frac{\partial \omega_N(x_2,y_1)}{\partial x} & \frac{\partial \omega_N(x_2,y_2)}{\partial x} & \cdots & \frac{\partial \omega_N(x_2,y_N)}{\partial x} \\
\vdots & \vdots & \ddots & \vdots \\
\frac{\partial \omega_N(x_N,y_1)}{\partial x} & \frac{\partial \omega_N(x_N,y_2)}{\partial x} & \cdots & \frac{\partial \omega_N(x_N,y_N)}{\partial x} 
\end{array}
\right),
\eeq
\no and so on.  The smallest eigenvalue of the Hessian matrix (\ref{hesse}) is given by
\beq
\label{eigenhesse}
\lambda_n(x,y) = \frac{1}{2} \left\lbrack 
\omega_{xx}^2 + \omega_{yy}^2 - {\left(
\omega_{xx}^2 + \omega_{yy}^2 - 2 \omega_{xx}\omega_{yy} + 4\omega_{xy}^2
\right)}^{1/2}
\right\rbrack,
\eeq
\no with corresponding eigenvector (non-normalized)
\beq
\label{eigenhesse1}
{\bf n}(x,y) = \left(
\begin{array}{c}
\omega_{xx}^2 - \omega_{yy}^2 - {\left(\omega_{xx}^2 + \omega_{yy}^2 - 2 \omega_{xx}\omega_{yy} + 4\omega_{xy}^2 \right)}^{1/2} \\
2 \omega_{xy}
\end{array}
\right).
\eeq

The ridges of the FTRN field are defined as the loci where the following conditions are fulfilled:
\beq
\label{ridges}
\nabla \omega_N(x,y) \cdot{\bf n} = 0, \qquad \lambda_n < 0,
\eeq
\no where
\beq
\label{gradomega}
\nabla \omega_N(x,y) = \left(
\begin{array}{c}
\frac{\partial \omega_N(x,y)}{\partial x} \\
\frac{\partial \omega_N(x,y)}{\partial y}
\end{array}
\right).
\eeq

\begin{figure}
\includegraphics[width=0.7\columnwidth,clip]{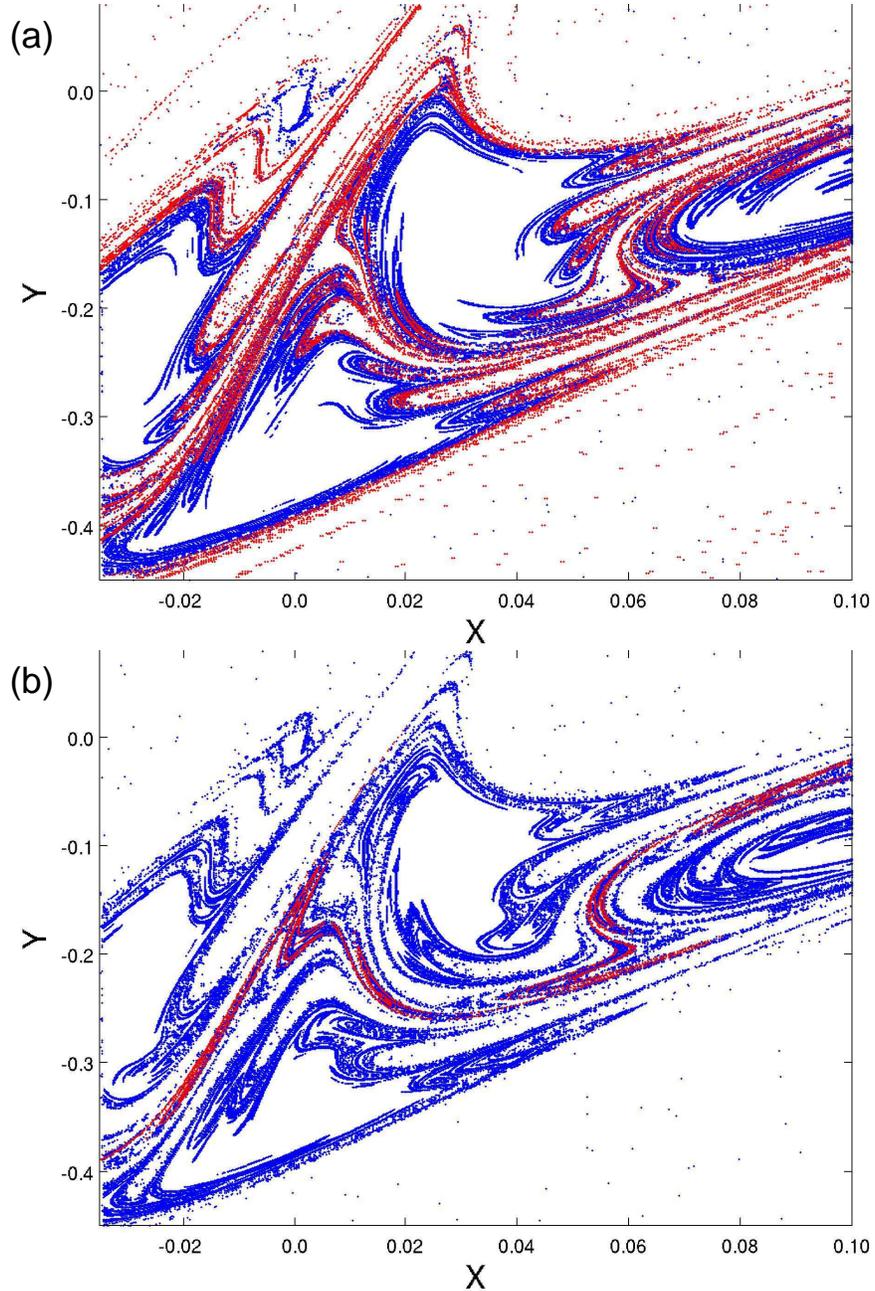}
\caption{\label{ridfig} (color online) Ridges of the finite-time rotation number field (blue) and invariant manifolds (red) for the standard nontwist map, with $a = 0.8063$ and (a) $b=0.80552$; (b) $b=0.80630$.}
\end{figure}

In Figs.~\ref{ridfig}(a) and (b) we plot the ridges of the FTRN  field corresponding to the cases of inter and intracrossings, respectively, so as to illustrate the usefulness of plotting the ridges for delineating escape channels. The ridges (in blue in Fig.~\ref{ridfig}) are plotted with the boundaries between positive and negative transport that were previously shown in Fig.~\ref{transfig} (in red), the latter indicating the escape channels for fast transport. In both cases the ridges act as the walls for the escape channels, in such a way that in the intracrossings [Fig.~\ref{ridfig}(b)] the ridges form effective transport barriers, whereas in the intercrossing situation depicted in Fig.~\ref{ridfig}(a) the barrier opens into a gateway for transport. Such features are responsible for the differences in transmissivity shown in Fig.~\ref{escapefig}.

\section{Conclusions}
\label{conclu}

One of the distinctive features of nontwist maps, and, in particular, of the standard nontwist map, is the capability of developing effective transport barriers, which hamper diffusion by means of  a trapping mechanism similar to that responsible for stickiness in Hamiltonian dynamical systems. These broken barriers are only effective on a timescale of the order of the experiment or the observation being conducted, for they are stickiness layers of chaotic behavior rather than true invariant tori.  That is,  one expects some transport to occur through these barriers, although on a timescale substantially larger than the typical duration of the experiment or numerical simulation. 

Such effective transport barriers can be considered as a type of dynamical structure, which are  interesting objects to focus attention on for analyzing transport in fluids and plasmas, inasmuch they separate regions of qualitatively dynamical behavior. For example, when observing the dispersion of pollutants in a marine environment,  effective transport barriers   may separate polluted regions from the surrounding waters on a time scale long enough to impede large scale spread of the pollutant. Since such structures change with time, one desires fast indicators that reveal the occurrence, if any, of any internal transport barriers.

For the standard nontwist map we investigated the parameter dependence of the transport through the broken shearless barrier. Upon varying a proper control parameter we identified the onset of high stickiness structures that give rise to the effective barrier near the broken shearless curve. Once this barrier had been formed, there are two qualitatively different phase space regions inside the barrier,  which were revealed by by the FTLE and FTRN indicators. We also observed the appearance of a period-three satellite island just after the formation of the effective transport barrier.

In the present analysis the computer time for the  FTRN is almost an order of magnitude less than that for the corresponding FTLE,  where we used the well-known method of \cite{wolf}.  We have shown in this paper the usefulness of both diagnostics on the simplest nontwist map presenting such internal transport barriers, but we claim that our results are directly applicable to problems of practical interest in geostrophic, atmospheric, and marine flows, as well as in the plasma confinement in tokamaks where the plasma has a nonmonotonic equilibrium profile.  


\section*{Acknowledgments}

This work was made possible by partial financial support of FAPESP, CNPq, CAPES, MCT/CNEN (Rede Nacional de Fus\~ao) and Funda\cao Arauc\'aria. P.J.M.\  was supported by U.S.~Dept.\ of Energy Contract \# DE-FG05-80ET-53088.

\end{document}